# Numerical Investigation of Pressure Losses and its Effect During Intake in a Steam Wankel Expander

Auronil MUKHERJEE[1], Satyanarayanan SESHADRI[1]*

[1]Indian Institute of Technology Madras, Department of Applied Mechanics,
Chennai-600036, Tamil Nadu, India
Tel : +91 44 2257 4072, e-mail : satya@iitm.ac.in

* Corresponding Author

## ABSTRACT

A small-scale volumetric Wankel steam expander has numerous advantages over other positive displacement machines as an expansion device. This is due to its high power to weight ratio, compactness, lower noise, vibration, and potentially lower specific cost making them a favourable choice over reciprocating expanders. The admission in the expander chamber occurs through rotary valves fed with a high-pressure steam supply from a boiler. Pressure drop of the steam is inevitable during the admission process due to the changing flow area of the rotary valve intake port during the admission process. The present study aims to investigate the magnitude of pressure losses of the steam during intake and its effect on the net power output of the expander over a range of rotational speed varying from 1200 to 3000 RPM. The thermodynamic analysis is carried out for the theoretical pressure-volume cycle of the expander using Python, which is then used for CFD analysis in Ansys Fluent 19.2®. Three dimensional models are developed for the flow domain stretching from the exit of the intake valve port to the port in the rotor housing at different rotor angles ranging from admission to cutoff. Numerical simulations are performed to investigate the flow dynamics and the associated pressure drop of steam during admission. The boundary conditions at different rotor angles are obtained from the thermodynamic model of the expander's theoretical pressure-volume plot, and the state point values were obtained from the REFPROP® database. Validation of the CFD models are carried out by comparing pressure drop values obtained through an analytical approach using correlations and expressions reported in literature. A grid independence test is performed to ensure a mesh-independent solution keeping residuals less than 10e-03. A thorough performance analysis of this expansion device is made and the loss in power output due to the intake pressure loss is calculated. These losses during admission change the pressure ratio across the actual expander designed for a given pressure ratio, leading to a reduced power output by a reasonable margin of 20 to 30%. It is observed that the percentage loss in power output increases with an increase in shaft speed.

## 1. INTRODUCTION

In recent years, the rapid consumption of fossil fuels has resulted in a slew of major environmental issues, including global warming, ozone layer depletion, acid rain, and air pollution. As global energy consumption and greenhouse gas emissions continue to rise, there is a pressing need to shift to sustainable energy sources, such as solar, biomass, and waste heat. A combined heat and power (CHP) system can be powered by solar thermal energy, biomass combustion heat, or waste heat. Several micro-CHP systems with organic Rankine cycles (ORC) appropriate for home applications (1-10kW) have been researched in recent years, using solar thermal, biomass-fired boilers, and waste heat resources. Small expansion devices are appealing for microgeneration applications because they allow renewable energy sources including geothermal, biomass, and solar energy. Volumetric expanders have higher efficiency than turbines at typical micro generating plant sizes.

In recent times, several researchers have looked into using a Wankel rotary engine as an expansion device. Since Felix Wankel discovered the rotary engine, some research has been conducted to improve its design and performance (Yamamoto, 1981a). Most research employed the Wankel engine as a combustion engine using various working fluids such as gasoline, methane, octane, hydrogen, diesel, and petrol, as in (Spreitzer *et al.*, 2015), while others used it as a compressor (Zhang et al., 2011) or an expander device (Francesconi and Antonelli, 2017). Antonelli *et al.* (2014a) compared simulation and experiments regarding supplied torque, working fluid mass flow rate, and indicated pressure for an Organic Rankine Cycle. The results were validated using compressed air. They used a lumped parameter numerical model to consider the losses due to leakage, friction and heat transfer. At 80 ºC, they achieved an isentropic efficiency of roughly 85% and a thermal cycle efficiency of 10% utilising pentane as the working fluid. Badr *et al.*, (1991) compared performance of Wankel expander to turbines, rotary vane, and helical-screw devices, demonstrating the advantages of employing the Wankel geometry as expansion devices, such as compactness, low vibration, noise, and low cost. Antonelli *et al.*, (2014b) theoretically investigated a small sized power plant using a steam Wankel expander operating with rotary intake and exit valves, utilising heat from renewable sources. When single stage were compared to multistage Wankel expanders, the results revealed a significant gain in thermal efficiency and a substantial drop in steam specific consumption of up to 25%. They also conducted experimental and numerical research on a small scale volumetric expansion device employing a variety of organic fluids, pressure ratios, and rotational speeds. The results revealed that employing R600-a at 1500 rpm and a saturated temperature of 110 ºC, an isentropic efficiency of 95% could be reached. Rosario (2005) used a Wankel expander for portable power applications to demonstrate the ability of producing electric power in the order of milliwatts, with an energy density better than batteries. Sadiq *et al.* (2017) conducted CFD simulations to develop a two stage Wankel expander and compare their performances to the single stage expander. They combined two Wankel expanders horizontally, with a larger one at the front and obtained a power output of 8.52kW compared with single-stage which gave 4.75kW power output at the same operating conditions.

Most of the studies done till now focused mainly on the Wankel expander's performance parameters in a wide range of operating conditions, operating it using different working fluid and modelling the leakage and thermal losses. Several investigations has also been done to observe the nature of flow dynamics and discharge coefficients through intake and exit valves . However, a detailed investigation of the estimation of net power output losses due to pressure losses across the intake manifold during admission through rotary valves were not presented. The saturated steam on exit from the intake valve port encounters a sudden expansion at the valve exit. It travels a finite distance to enter the expander chamber through the port in the rotor housing. Pressure losses are inevitable during this process. These make it worthwhile to investigate the magnitude of the losses and its effect on the net power output of the expander for a wide range of RPMs. This is the focus of the present study, for a Wankel expander prototype established in the Energy and Emissions Lab, IIT Madras (Gopal and Seshadri, 2022). In the following sections, a thermodynamic state point analysis of the expander is performed in Python coupled with the NIST-REFPROP® database to visualize the trend of pressure volume variation inside the expander chamber and verify with reported literature (Gopal and Seshadri, 2022). Subsequently, a thorough numerical investigation estimates the pressure loss variation with the rotor angle during the entire admission duration at five different RPMs ranging from 1200 to 3000. Leakage and thermal losses are neglected and the numerical results are validated using an analytical approach using pressure drop correlations and expressions reported in literatures by (Idelchik, 1986) and (Duan *et al.*, 2012). Consequently, the present work provides a preliminary investigation on the effect of pressure drop across the intake manifold on the power output of a volumetric expansion device and its variation with the shaft speed.

## 2. METHODOLOGY

### 2.1 Wankel Expansion Device Geometry
The Wankel expander prototype consists of the static housing , two moving components, the triangular rotor and the eccentric output shaft and two rotary intake and exhaust valve. An internal and external spur gear governs the rotor's motion. The external gear is fixed to the side housing, and the internal gear is fixed within the rotor. This keeps the tip of the rotor in contact with the housing (Yamamoto, 1981b). The radius *R* of the rotor and the eccentricity *e* of the output shaft determine the geometry of the rotor housing and flanks. Both are the key design parameters for the Wankel expander as demonstrated in Fig. 1.Point *A* and *O* are the housing and rotor centre respectively. The length *OA* and *OB* in Fig 1 denotes the eccentricity and rotor radius respectively. The rotor rotates around its centre and undergoes translation motion along the eccentric shaft of radius *e*. One revolution of the rotor around its centre corresponds to three revolution of the eccentric shaft and the rotary valves.

The housing's parametric equations are as follows:

$$x_h = e \cos 3\theta + r \cos \theta \tag{1}$$

$$y_h = e \sin 3\theta + r \sin \theta \tag{2}$$

The equations for the shape of the rotor are as follows:

$$x_r = r \cos 2\upsilon + \frac{3e^2}{2R}(\cos 8\upsilon - \cos 4\upsilon) \pm e\left(1 - \frac{9e^2}{R^2}\sin^2 3\upsilon\right)^{\frac{1}{2}}(\cos 5\upsilon + \cos\upsilon)$$

$$y_r = r \sin 2\upsilon + \frac{3e^2}{2R}(\sin 8\upsilon - \sin 4\upsilon) \pm e\left(1 - \frac{9e^2}{R^2}\sin^2 3\upsilon\right)^{\frac{1}{2}}(\cos 5\upsilon + \cos\upsilon)$$

In expression (3) and (4) the intervals of $\upsilon$ are as follows:

$$\upsilon = \left[\frac{\pi}{2}, \frac{5\pi}{6}\right], \left[\frac{11\pi}{6}, \frac{13\pi}{6}\right], \left[\frac{19\pi}{6}, \frac{21\pi}{6}\right]$$

**Table 1:** Wankel Expander Design Parameters

| Parameter | Value | Unit |
|---|---|---|
| R | 80.0 | mm |
| e | 10.5 | mm |
| b | 71.0 | mm |

### 2.2 Thermodynamic Analysis

The Wankel Expander is inherently deigned for a specified pressure ratio. The intake and exhaust pressure limits are kept considering the boiler and condenser pressures of a Rankine Cycle system, in which the Wankel has to serve as a expansion device. If $P_b$ and $P_c$ are boiler and condenser pressures respectively, the pressure ratio of the expander is $r_p$:

$$r_p = \frac{P_b}{P_c} \quad (5)$$

A schematic of a theoretical pressure-volume variation for the expander is depicted in Figure 2. This cycle occurs twice for a single rotation of the triangular rotor, thereby delivering six power stroke in one full rotation of the triangular rotor (Badr *et al.*, 1991). The ratio between valve shaft speed and rotor shaft speed is kept 1:1 as reported in (Gopal and Seshadri, 2022). The operating parameters of the expansion device are listed in Table 2.

The cylindrical intake rotary valve port remains fully closed when the rotor is at the clearance volume. The port starts to open as the process of steam admission begins, reaches a fully opened stage and gets fully closed at the cut off. The expansion process begins and continues until the expander chamber reaches the maximum volume. The steam is in saturated vapour state during admission and in the liquid-vapour mixture state during the expansion process. The steam quality at the maximum chamber volume is obtained by equating the entropy at the beginning and end of expansion. The specific volume at the end of expansion is obtained using the REFPROP database. The maximum and minimum volume of the Wankel chamber can be calculated by taking the extremum of the expression (6). The isentropic index is calculated using the adiabatic law at the beginning and end of the expansion process. Neglecting leakage loss during the admission and expansion process and using mass conservation, the mass of steam expanded in the chamber is obtained which gives the mass of steam intake in a single cycle.

**Table 2:** Wankel Expander Operating Parameters

| Parameter | Range/Value | Unit |
|---|---|---|
| $P_b$ | 10 | bar |
| $T_{in}$ | 453 | K |
| $P_c$ | 3 | bar |
| N | 1200-3000 | RPM |

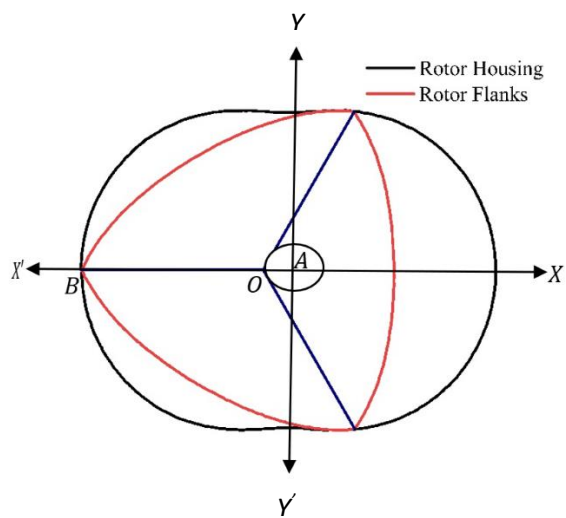
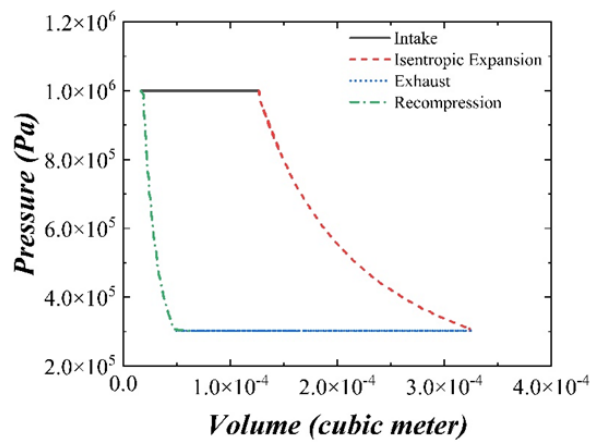

**Figure 1**: Schematic of Wankel Geometry.

**Figure 2**: Ideal pressure volume variation in a single chamber.

### .2.3 Measurement of Volume and Mass Flow Rate

The rotation of the triangular rotor inside the housing results in a continuous change in the volume of the working chamber of the expander throughout its operation. This volume is calculated by multiplying the side area contained by a rotor contour side and the inner surface of the epitrochoid by the rotor housing width. The volume of the working chamber is a function of the expander design parameters and the shaft angle. It is expressed as (Yamamoto, 1981c):

$$V(\theta) = \frac{\pi}{3}e^2 + eR\left(2\left(1 - \frac{9e^2}{R^2}\right)^{\frac{1}{2}} - \frac{3\sqrt{3}}{2}\sin\left(\frac{2\theta}{3} + \frac{\pi}{2}\right)\right) + \left(\frac{2}{9}R^2 + 4e^2\right)\sin^{-1}\left(\frac{3e}{R}\right) \quad (6)$$

The volume flow rate of steam in the chamber is obtained by evaluating the derivative of expression (6) and multiplying with the shaft speed. The corresponding mass flow rate during admission till cut-off is obtained by the product of the volume flow rate with the density of the admitted steam. The mathematical approach is shown below:

$$\frac{dV}{dt} = \frac{\partial V}{\partial \theta} * \frac{\partial \theta}{\partial t} = \omega \frac{\partial V}{\partial \theta} \quad (7)$$

$$\frac{dm}{dt} = \rho_{ad}\omega \frac{\partial V}{\partial \theta} \quad (8)$$

The volume of the working chamber at steam cut off for the design pressure ratio of the Wankel expander, is obtained based on aforementioned thermodynamic analysis. Subsequently, the variation of mass flow rate with rotor angle at different rotational speeds is studied during the intake duration. Figure 3 represents the variation of the mass flow rate inside the chamber volume. The magnitude of the flow rate reaches a maxima before the closing of the intake valve port and its magnitude increases with increase in shaft speed.

The next section of the paper investigates pressure losses across the intake manifold for the entire admission duration using CFD. Numerical simulations of the variation of the losses at different rotor angle instants and its trend with increasing shaft speed are presented. The consequences of these losses on the net power output of the expander is investigated.

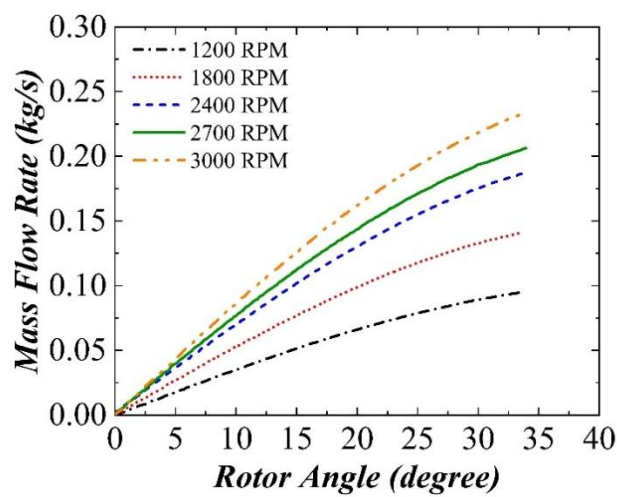

**Figure 3:** Variation of Mass Flow Rate with rotational speed.

### 2.4 Computational Fluid Dynamics (CFD)
Computational fluid dynamics is an efficient method for simulating the behaviour of thermo-fluids in a system. It's a useful tool for analysing, optimising, and verifying design performance before constructing prototypes and conducting physical tests. Finite volume based commercial package Ansys Fluent 19.2® is used in the present study for 3D modelling, meshing, numerical simulation and necessary postprocessing.

#### 2.4.1 Three-dimensional computational detail
The computational domain in the present study consists of the intake manifold of the Wankel Expander extending from the exit of the intake valve port to the port present in the rotor housing. Three dimensional CAD models are developed in Ansys Design modular for each angular instants of the rotor from admission to cut-off. The inlet area of the domain is adjusted depending upon the rotor angle of the expansion device. Figure 4 depicts a schematic of the flow domain at a rotor angle of 9º. The black arrow denotes the direction of the flow stream. A schematic of the intake port geometry is shown in Figure 5. The rotation of the intake rotary valve during admission causes the port to open and close along its width. The domain is discretized using a hexahedral meshing technique that ensures the mesh is locally structured and can maintain orthogonal grids in the wall normal direction. This is due to the higher precision of the hexahedral components which allows the angle between faces to be retained near to 90 degrees. A schematic of the meshed geometry is depicted in Figure 6.

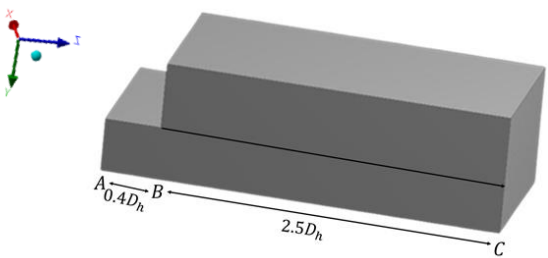
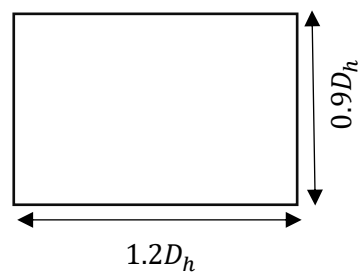

**Figure 4:** Schematic of the computational domain.     **Figure 5:** Schematic of the valve port geometry.

#### 2.4.2 Numerical Method and Governing Equations
The governing equations which are used for the fluid flow modelling based on the mass, momentum and energy conservation are mentioned in equation (9), (10) and (11) in the vector form. The flow is incompressible with no slip on the wall boundary and adiabatic condition. The atmospheric pressure and ambient temperature are taken 1.013 bare and 300K respectively, as per ANSYS Fluent 19.2® user manual. The transient pressure-based solver is selected to ensure a time-dependent flow solution at each rotor angular instants. Turbulence is simulated using the renormalization group k-epsilon (RNG) model, and numerical simulation is done using the 'coupled' pressure-velocity approach.
The three dimensional Continuity, Momentum and Energy equations for incompressible flow with constant fluid properties are as follows:

$$\nabla \cdot \vec{U} = 0 \tag{9}$$

$$\rho \frac{D\vec{U}}{Dt} = \rho \vec{g} - \nabla P + \mu \nabla^2 \vec{U} \tag{10}$$

$$\rho c_p \frac{DT}{Dt} = \rho \dot{q} + k \nabla^2 T + \varphi \tag{11}$$

The transport equations for the RNG k-epsilon Model are:

$$\frac{\partial}{\partial t}(\rho k) + \frac{\partial}{\partial x_i}(\rho k u_i) = \frac{\partial}{\partial x_j}\left(\alpha_k \mu_{eff} \frac{\partial k}{\partial x_j}\right) + G_k + G_b + \rho \varepsilon - Y_M + S_k \tag{12}$$

$$\frac{\partial}{\partial t}(\rho \varepsilon) + \frac{\partial}{\partial x_i}(\rho \varepsilon u_i) = \frac{\partial}{\partial x_j}\left(\alpha_\varepsilon \mu_{eff} \frac{\partial \epsilon}{\partial x_j}\right) + C_{1\varepsilon} \frac{\varepsilon}{k}(G_k + C_{3\varepsilon} G_b) - C_{2\varepsilon} \rho \frac{\epsilon^2}{k} - R_\varepsilon + S_\varepsilon \tag{13}$$

In the above two equations, the generation of turbulence kinetic energy due to the mean velocity gradients is denoted by $G_k$ and $G_b$ is the generation of turbulence kinetic energy due to buoyancy. $Y_M$ represents the contribution of the fluctuating dilatation in compressible turbulence to the overall dissipation rate. $\alpha_k$ and $\alpha_\varepsilon$ are the inverse effective Prandtl numbers for $k$ and $\varepsilon$ respectively. The user defined source terms are denoted by $S_k$ and $S_\varepsilon$.

### 2.4.3 Initial and Boundary Conditions
The steam from the boiler enters the intake manifold of the Wankel Expander at a constant pressure through the port present in the rotary intake valve as illustrated in Figure 5. This serves as the inlet of the flow domain. The inlet boundary condition are as follows:

$$P = P_b, \quad T = T_{in} \tag{14}$$

Mass flow rate boundary condition is imposed at the outlet of the flow domain. The magnitude of the mass flow rate, which is a function of the shaft angle of the expander is adjusted depending upon the inlet area of the flow domain using equation (6), (7) and (8). The no-slip and adiabatic boundary conditions is applied to the manifold's upper and lower wall.

### 2.4.4 Grid Independence Study
A grid sensitivity analysis is conducted to determine the optimum mesh size for the numerical simulations. This is done by calculating the pressure loss at a rotor angle of 36 degree just before the cut-off, at a rotational speed of 1200 RPM at five different element sizes.

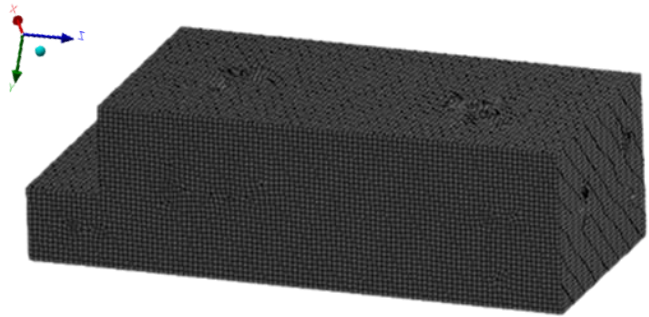

**Figure 6**: Computational domain of the intake manifold using hexahedral meshing.

The error is determined on the basis of the finest grid and the residual is found to be less than 10e-03. Consequently, this mesh size is found to be optimum for all other angular instants and shaft speeds. Figure 7 depicts a schematic of the variation of the pressure losses at five different mesh sizes in the aforementioned conditions.

### 2.4.5 Model Validation
The intake manifold of the expander stretches from the rectangular shaped port of the intake valve to the port on the rotor housing. The numerical results and the flow physics are validated using an analytical approach by calculating the net head loss from inlet to outlet of the flow domain at different rotor angles , thereby calculating the pressure drop. The steam entering the domain suffers a frictional head loss in the region AB and BC as shown in Figure 4 and a head loss due to the sudden expansion at point B. The head losses in region AB and BC are calculated using the modified Colebrook-White equation proposed by Duan *et al.* (2012) and the loss due to sudden expansion at point B is calculated using the expression of Idelchik (2008) which is derived for practical scenarios for nonuniform velocity distribution at high Reynolds number. The magnitude of the losses at different rotor angular instants are calculated for a rotational speed of 2400 RPM. The analytical and numerical results follow similar trends and have a reasonable match with a root mean squared error magnitude of 5%. A comparative schematic of the pressure loss variation with rotor angle is shown in Figure 8.

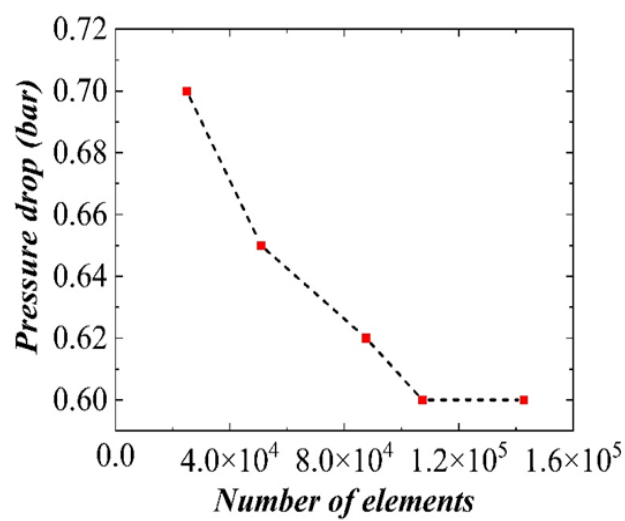
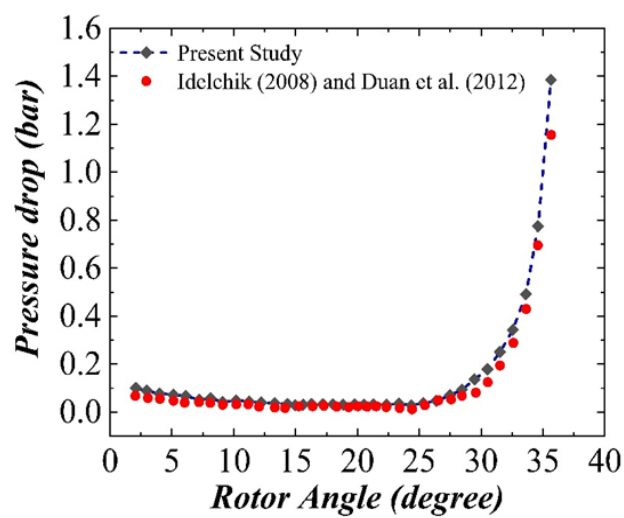

**Figure 7:** Schematic of the grid independence test for different mesh sizes.

**Figure 8:** Numerical variation of pressure loss during intake compared to literature.

## 3. RESULTS AND DISCUSSIONS

### 3.1 Pressure Drop Variations during Intake
The saturated steam enters the intake manifold at the expander clearance volume and continues until cut-off. The steam encounters frictional losses during its motion before it enters the expander chamber through the port in the rotor housing. The variation of the pressure losses during the entire admission duration is illustrated for five different RPMs ranging from 1200 to 3000 in Figures 9 and 10. The magnitude of the loss decreases with an increase in rotor angle during the opening duration of the port and reaches a minimum value when the port is fully opened. This is due to the increase in the inlet area of the manifold during the opening duration. The pressure drop starts to increase during the closing duration of the port rapidly. This is due to the reduction in the inlet flow area and the increase in mass flow rate with rotor angle, as depicted in Figure 3. The magnitude of the pressure losses also shows an increasing trend with an increase in shaft speed. It is observed that the loss varies from 10 to 30% of the intake pressure in the operating range of RPMs. The expansion device is inherently designed for a fixed pressure ratio. These pressure losses during admission change the design pressure ratio of the expander, which has direct consequences on the net power output of the expander. A schematic of the pressure contours on the intake manifold at two different rotor angular instants at 3000 RPM is depicted in Figures 11 and 12. The majority of pressure losses happen at the valve port exit due to the sudden expansion in the cross-sectional area. After exit, there is some degree of pressure recovery of the steam before it finally enters the chamber volume through the rotor housing port, which is the corresponding outlet of the intake

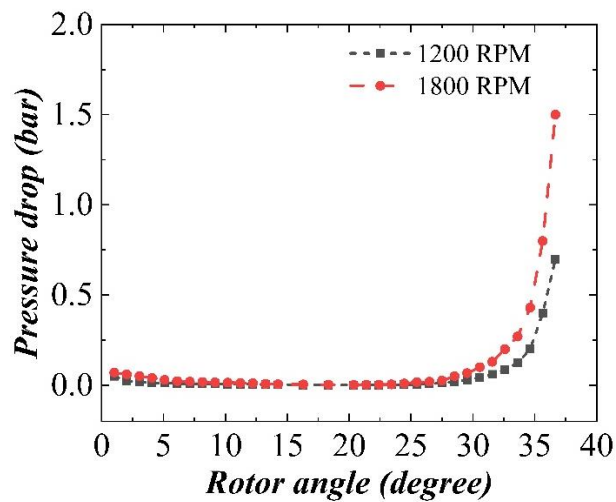
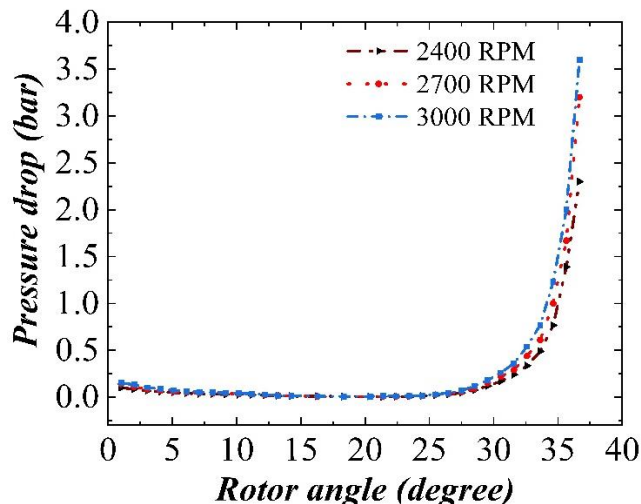

**Figure 9:** Pressure drop variation with rotor angle for 1200 and 1800 RPM.

**Figure 10:** Pressure drop variation with rotor angle for 2400, 2700 and 3000 RPM.

manifold.

### 3.2 Effect of Pressure drop on Net Power Output

The loss in pressure in the intake manifold decreases the design pressure ratio of the expansion device, which has direct consequences on the net power output of the expander. This, in turn, reduces the expansion stroke of the Wankel expander, which leads to a loss in power output by a reasonable margin of 15 to 35 %. The percentage loss in power output increases with an increase in shaft speed, as illustrated in Figure 14. A revised pressure-volume plot incorporating the intake losses is depicted in Figure 13 at a shaft speed of 2700 RPM. The green shaded region illustrates the amount of work potential loss with respect to the ideal work done by the expander in a single cycle. The work done in a single cycle is obtained by calculating the enclosed area of the pressure-volume curve using the polyarea function in MATLAB® as demonstrated in equation (15-18). The net power output of the expander is found out using the approach shown below :

$$W_{single\ cycle} = A_{enclosed}(P,V) \tag{15}$$

$$A_{enclosed} = polyarea(P,V) \tag{16}$$

$$P_{out,single\ cycle} = 6 \times (A_{enclosed}) \tag{17}$$

$$P_{net} = P_{out,single\ cycle} \times \frac{N}{3} \tag{18}$$

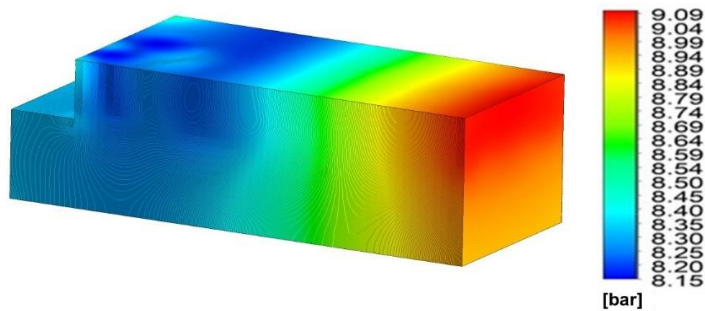
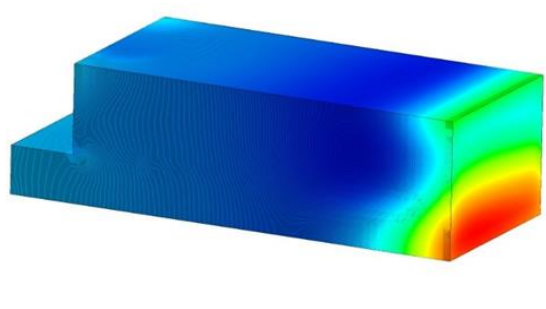

**Figure 11:** Pressure contour at 28.5 degree Rotor angle, 3000 RPM.

**Figure 12:** Pressure contour at 34 degree Rotor angle, 3000 RPM.

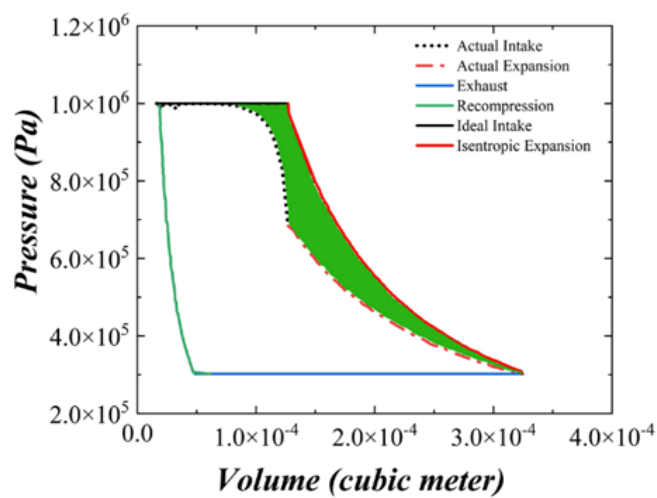
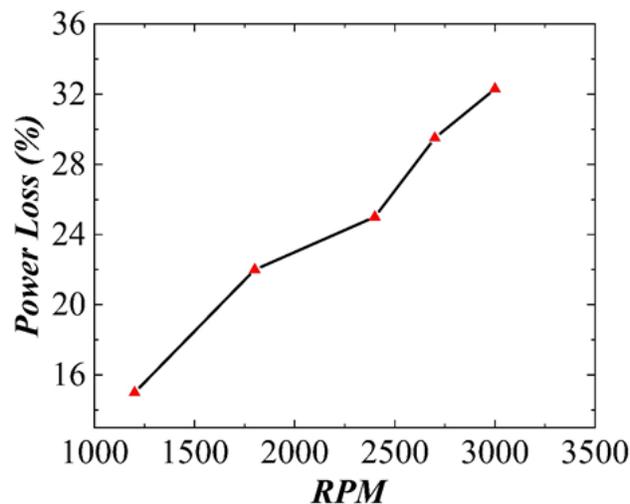

**Figure 13:** Pressure volume plot considering intake losses at 2700 RPM.

**Figure 14**: Variation of percentage power loss with RPM.

## 4. CONCLUSIONS

A preliminary investigation is conducted using CFD simulation to find the variation of pressure loss across the intake manifold of a steam Wankel expander operating with rotary valves and its effect on the net power output of the expansion device. Thermodynamic modelling of the expander is conducted based on its design pressure ratio and geometric parameters. CFD results revealed that pressure loss across the intake manifold decreases during the opening of the valve shaft port and shows a rapid increase during the closing duration. This phenomenon occurs due to increment and decrement in the inlet flow area due to opening and closing of the intake port with the shaft rotation. The mass flow rate increases continuously from the beginning of intake till cut-off, which results in higher pressure losses during the closing duration of the port. The magnitude of these losses increases with an increase in the shaft speed of the expander. The expander is inherently designed for a given pressure ratio. Pressure losses across the intake manifold reduce the design pressure ratio, which reduces the net power output of the expander by a reasonable margin of 15 to 35 percent. It is also observed that the magnitude of power loss increases with an increase in shaft speed. While it is necessary to operate the expander at a higher shaft speed to attain more power output and reduce leakage losses, these pressure losses across the intake manifold act as a counteracting phenomenon that diminishes the work potential of the volumetric expansion device.

## NOMENCLATURE

| | | |
|---|---|---|
| $A$ | area | mm$^2$ |
| $b$ | rotor Width | mm |
| $c_p$ | specific heat at constant pressure | J/kgK |
| $D_h$ | hydraulic diameter | mm |
| $e$ | eccentricity | mm |
| $N$ | shaft speed per minute | RPM |
| $P$ | pressure | N/m$^2$ |
| $R$ | rotor radius | mm |
| $T$ | temperature | K |
| $\vec{U}$ | velocity vector | ms$^{-1}$ |
| $V$ | Volume | mm$^3$ |
| $W_{single\ cycle}$ | work done in single cycle | J |
| $\dot{q}$ | rate of heat generation per unit volume | W/m$^3$ |
| $\rho$ | density | kg/m$^3$ |
| $k$ | Thermal conductivity | Wm$^{-1}$ K$^{-1}$ |
| $\mu$ | dynamic viscosity | Ns/m$^2$ |

| $\theta$ | shaft angle | ° |
| $P_{net}$ | net power output | W |

**Subscripts**

| b | boiler |
| c | condenser |
| ad | admission |
| in | inlet |

# ACKNOWLEDGEMENT

The authors are thankful the to PG Senapathy Computing Resources at IIT Madras for providing the essential computing resources for the present study.